\documentclass[prd,showpacs,showkeys,nofootinbib,twocolumn]{revtex4-1}

\begin{document}

\title{Equivalence principle in scalar-tensor gravity}

\author{Dirk Puetzfeld}
\email{dirk.puetzfeld@zarm.uni-bremen.de}
\homepage{http://puetzfeld.org}
\affiliation{ZARM, University of Bremen, Am Fallturm, 28359 Bremen, Germany} 

\author{Yuri N. Obukhov}
\email{obukhov@ibrae.ac.ru}
\affiliation{Theoretical Physics Laboratory, Nuclear Safety Institute, 
Russian Academy of Sciences, B.Tulskaya 52, 115191 Moscow, Russia} 

\date{ \today}

\begin{abstract}
We present a direct confirmation of the validity of the equivalence principle for unstructured test bodies in scalar tensor gravity. Our analysis is complementary to previous approaches and valid for a large class of scalar-tensor theories of gravitation. A covariant approach is used to derive the equations of motion in a systematic way and allows for the experimental test of scalar-tensor theories by means of extended test bodies.
\end{abstract}


\maketitle


\section{Introduction}

Since the famous tower observations by Galileo, the independence of the dynamics on the mass of a test body in a gravitational field has been verified in numerous physical experiments \cite{Will:1993,Will:2014}. This striking property of test matter, later on termed ``equivalence principle'', was put by Einstein at the foundation of General Relativity theory, further generalizations also encompass classical and quantum extensions of General Relativity \cite{Ni:1977,Ni:1983,Hehl:1991,Lemke:1994,Lammerzahl:1996,Lammerzahl:2003}. Tests of the equivalence principle are therefore of fundamental importance for relativistic gravitational theories. 

Scalar-tensor theories are considered to be close and viable generalizations of Einstein's general relativity theory. Since their early introduction in \cite{Jordan:1955,Jordan:1959,Thiry:1951} they attracted a lot of attention in the literature, in particular after the works of Brans and Dicke \cite{Brans:Dicke:1961,Brans:1962:1,Brans:1962:2,Dicke:1962:1,Dicke:1964} in which the scalar field was interpreted as a variable gravitational coupling -- for an overview of the history and results of scalar-tensor theories see \cite{Fujii:Maeda:2003,Brans:2005,Goenner:2012,Sotiriou:2014}.

Despite the long history of scalar-tensor theories, surprisingly little attention was paid to the investigation of motion of extended test bodies in such theories. After some preliminary discussions \cite{Brans:1962:2,Bergmann:1968,Wagoner:1970}, the dynamics of compact bodies was thoroughly studied in \cite{Damour:etal:1992} in the framework of the post-Newtonian formalism.

Here we present a multipolar derivation of the dynamics of extended test bodies. In particular, we demonstrate the validity of the equivalence principle for unstructured test bodies for a very large class of scalar-tensor theories. We use the notion of ``test bodies'' along the lines of Infeld and Plebanski \cite{Infeld:1960}, who distinguished three kinds of equations of motion. According to them, the equations of the first kind describe the motion of a test body (particle) under the action of a given external field which does not depend on the dynamics of the test body. The equations of the second and of the third kinds take into account the back reaction, so that a body moves in a field that depends on the mass and the motion of an extended body. Here we deal with the equations of motion of the first kind. 

We study the class of scalar-tensor gravity models which is fairly general. In the Brans-Dicke-(Jordan-Thiry) theory the gravitational and scalar fields are universally coupled to matter of all types. However, in \cite{Damour:1990} this postulate was relaxed, and while preserving the universality of the gravitational coupling, it was assumed that scalar fields may couple differently to visible (ordinary) and to invisible (dark) matter. Later this idea was developed into a scalar chameleon theory \cite{Khoury:PRL:2004,Khoury:PRD:2004}. 

Our analysis is complementary to the ones in \cite{Hui:2009,Hui:2010} and \cite{Gralla:2010,Gralla:2013}. It clearly demonstrates, that future experiments to test scalar-tensor gravity should either make use of structured test bodies or heavy bodies.  

\section{Conservation of energy and momentum in scalar-tensor theory}

The conservation law of the energy-momentum tensor underlies the analysis of the equations of motion of the first kind. We suppose that the dynamics of matter that constitutes a test body is described by the matter Lagrangian ${L}_{\rm mat}$. The latter depends on the material fields (and their derivatives) which interact with the spacetime metric $g_{ij}$ and with a multiplet of scalar fields $\varphi^A$ (capital indices $A,B,C = 1,\dots, N$ label the components of the multiplet). 

The metrical energy-momentum tensor of matter is constructed as usual via $\sqrt{-g}t_{ij} := 2 \partial (\sqrt{-g}{L}_{\rm mat})/\partial g^{ij}$. From the Noether theorem for diffeomorphism invariant models one finds the generalized conservation law
\begin{equation}\label{cons} 
\nabla_jt^{kj} = \left(\alpha\,t^{kj} - \beta\,g^{kj}g_{mn}t^{mn}\right){\frac 1F}\partial_jF = - \,V_{ij}{}^kt^{ij}.
\end{equation}
Here, quite generally, $F = F(\varphi^A)$ and we introduced
\begin{equation}\label{AZ}
V_{ij}{}^k = -\,\alpha\,A_j\delta_i^k + \beta\,g_{ij}A^k,\qquad A_i := \partial_i\log F.
\end{equation}
Different scalar-tensor models are characterized by specific values of the constants $\alpha$ and $\beta$. For example, in standard Brans-Dicke theory $\alpha = 4$ and $\beta = 1$, see \cite{Obukhov:Puetzfeld:2014:2}.
In chameleon theory \cite{Khoury:PRL:2004,Khoury:PRD:2004}, the values of $\alpha$ and $\beta$ depend on the matter type, see Appendix \ref{CH}. 

It is worthwhile to stress that the important feature of the multipolar approach, which we pursue in the next section to derive the dynamics of test bodies, is that the result does not depend on the explicit form of the Lagrangians for the gravitational and the scalar fields.

\section{Equations of motion}

We derive equations of motion using the Math\-isson-Papapetrou-Dixon \cite{Mathisson:1937,Papapetrou:1951:3,Dixon:1964,Dixon:1974} approach by integrating the conservation law (\ref{cons}). This can be done in a most convenient way with the help of the geodesic expansion technique of Synge \cite{Synge:1960}. Denoting the world function by $\sigma$ and the parallel propagator by $g^{y}{}_{x}$, we introduce integrated moments to an arbitrary multipolar order $n=0,1,2,\dots$ by:
\begin{eqnarray}
p^{y_1 \dots y_n y_0}&:=& (-1)^n  \int\limits_{\Sigma(s)}\sigma^{y_1} \cdots \sigma^{y_n} g^{y_0}{}_{x_0}\sqrt{-g}t^{x_0 x_1} d \Sigma_{x_1}, \nonumber \\ \label{p_moments_def} \\
k^{y_2 \dots y_{n+1} y_0 y_1}&:=& (-1)^{n}  \int\limits_{\Sigma(s)} \sigma^{y_1} \cdots \sigma^{y_n} g^{y_0}{}_{x_0}g^{y_1}{}_{x_1} \nonumber \\ 
 && \sqrt{-g}t^{x_0 x_1} w^{x_2} d \Sigma_{x_2}.\label{k_moments_def}
\end{eqnarray}
Here we use a condensed notation so that $y_{n}$ denotes indices at the point $y$. The point $y$ we associate with the world-line $y(s)$ of an extended test body, parametrized by the proper time $s$. As usual, the integrals are performed over spatial hypersurfaces $\Sigma(s)$.

\subsection{Pole-dipole equations of motion}

In the pole-dipole approximation, an extended body is characterized by the multipole moments $p^a, p^{ab}, k^{ab}, k^{abc}$. Using the general multipolar scheme \cite{Puetzfeld:Obukhov:2014} we derive the equations of motion for these moments:
\begin{eqnarray}
0 &=& k^{(a|c|b)} - v^{(a} p^{b)c}, \label{eom_2_n_2_ST} \\
\frac{D}{ds} p^{ab} &=&  k^{ba} - v^a p^b  - V_{dc}{}^b k^{acd},\label{eom_2_n_1_ST}\\ 
\frac{D}{ds} p^{a} &=& - V_{cb}{}^a k^{bc} - V_{dc}{}^a{}_{;b} k^{bcd} 
-\frac{1}{2} R^a{}_{cdb} \left(k^{bcd} + v^d p^{bc} \right). \nonumber \\ \label{eom_2_n_0_ST} 
\end{eqnarray}
Here $v^a:=dy^a/ds$ denotes the normalized four-velocity of a body. Since $k^{a[bc]} = 0$, we can solve (\ref{eom_2_n_2_ST}) to find explicitly
\begin{eqnarray}
k^{abc} = v^a p^{cb} + v^cp^{[ab]} + v^bp^{[ac]} + v^ap^{[bc]}.\label{kabcST}
\end{eqnarray}
Plugging this into (\ref{eom_2_n_1_ST}) and (\ref{eom_2_n_0_ST}) and taking into account (\ref{AZ}), we obtain the generalized Mathisson-Papapetrou-Dixon system
\begin{eqnarray}
{\frac {D{\cal P}^a}{ds}} &=& {\frac 12}R^a{}_{bcd}v^b{\cal J}^{cd} - \beta\xi f^a - \beta\xi^b\nabla_bf^a, \label{DPtotST}\\
{\frac {D{\cal J}^{ab}}{ds}} &=& -\,2v^{[a}{\cal P}^{b]} - 2\beta\xi^{[a}f^{b]}.\label{DJtotST}
\end{eqnarray}
Here $f^a := F^{-\alpha}A^a$, and following \cite{Puetzfeld:Obukhov:2013:3,Puetzfeld:Obukhov:2014,Obukhov:Puetzfeld:2014:2}, we introduce the generalized total energy--momentum 4-vector and the generalized total angular momentum by
\begin{eqnarray}
{\cal P}^a &:=& F^{-\alpha}p^a + p^{ba}\nabla_bF^{-\alpha},\label{PtotST}\\
{\cal J}^{ab} &:=& F^{-\alpha}L^{ab}.\label{JtotST}
\end{eqnarray}
The orbital angular moment is defined by $L^{ab} := 2p^{[ab]}$, and we denoted
\begin{eqnarray}
\xi^a := g_{bc}k^{abc},\qquad \xi := g_{ab}k^{ab}.\label{xika}
\end{eqnarray} 

\subsection{Monopolar equations of motion}

At the monopolar order, the only nontrivial moments are $p^a$, and $k^{ab}$. The system (\ref{eom_2_n_2_ST})-(\ref{eom_2_n_0_ST}) then reduces to
\begin{eqnarray}
0 &=& k^{ba} - v^a p^b,\label{Mono1ST}\\
{\frac {Dp^a}{ds}} &=& -\,V_{cb}{}^ak^{bc}.\label{Mono2ST}
\end{eqnarray}
Making use of $k^{[ab]} = 0$, the first equation yields $v^{[a} p^{b]} = 0$, hence we have 
\begin{equation}
p^a = Mv^a\qquad\Longrightarrow\qquad \xi = M,\label{pMv}
\end{equation}
with the mass $M := v^ap_a$. Substituting (\ref{Mono1ST}) and (\ref{pMv}) into (\ref{Mono2ST}) we find
\begin{equation}\label{Mono3ST}
{\frac {D(Mv^a)}{ds}} = \alpha Mv^a {\frac 1F}{\frac {dF}{ds}} -\,\beta M{\frac 1F}\nabla^aF.
\end{equation}
Contracting this with $v_a$, we derive 
\begin{equation}
{\frac {dM}{ds}} = M(\alpha - \beta){\frac 1F}{\frac {dF}{ds}},\label{MFdot}
\end{equation}
and with the help of this we write (\ref{Mono3ST}) in the final form
\begin{equation}\label{MonoF}
{\frac {Dv^a}{ds}} = -\,\beta(g^{ab} - v^av^b){\frac {\nabla_bF}{F}}. 
\end{equation}
Quite remarkably, we thus find that the dynamics of an extended test body in the monopole approximation is independent of the body's mass. In case of a trivial coupling function $F$, equation (\ref{MonoF}) reproduces the well known general relativistic result. 

Interestingly, the mass of a body is not constant: its dynamics is described by (\ref{MFdot}): we can solve this differential equation to find explicitly the dependence of mass on the scalar function: $M = F^{\alpha - \beta}M_0$ with $M_0=$ const.

\section{Conclusions}

Our main result is the system (\ref{DPtotST})-(\ref{DJtotST}) that describes the dynamics of extended test bodies in scalar-tensor gravity. This is a direct generalization of the classic general-relativistic Mathisson-Papapetrou-Dixon result. The integration of these equations of motion (although a nontrivial task), should form the basis for local systematic tests of scalar-tensor gravity by means of spinning extended test bodies. 

In the monopolar case, our analysis revealed a surprisingly simple equation of motion (\ref{MonoF}). In contrast to geodesic motion in General Relativity, freely falling massive test bodies in scalar-tensor gravity experience an additional force, determined by the new scalar degrees of freedom encoded in the function $F$. The simplicity of (\ref{MonoF}) makes this equation an ideal candidate for the use in combination with free fall experiments. 

We stress that our method is complementary to the ones used in \cite{Hui:2009,Hui:2010,Gralla:2010,Gralla:2013}. In contrast to other methods it does not require the use of the full field equations due to its limitation to the test body case, and therefore benefits from a certain simplicity. In particular it allows for a straightforward generalization to other gravity theories \cite{Puetzfeld:Obukhov:2014}, which significantly go beyond the framework of scalar-tensor theories. Again we stress, that our results do not depend on the explicit form of the Lagrangians for the gravitational and the scalar fields.

A remarkable feature of (\ref{MonoF}) is the prediction that all massive test bodies move in the same way, independently of their mass. We thus demonstrate the validity of the equivalence principle in scalar-tensor gravity. Namely, in accordance with the weak equivalence principle, the trajectory of a test particle depends only on the initial position and velocity of the body, but not on its mass or internal structure. It is worthwhile to note that such a conclusion is valid for a wide class of models, including the generalized Brans-Dicke theory (with $\beta = 1$) and also for the chameleon theory (with $\beta\neq 1$).

This result is consistent with the previous independent analysis \cite{Hui:2009,Hui:2010}, which has shown that the total scalar charge of a body is equal to its mass when the scalar field self-interactions are neglected. The latter is in agreement with the test body assumption that underlies the Mathisson-Papapetrou-Dixon approach, yielding equations of motion of the first kind. When one goes beyond the test body approximation, however, the scalar charge is no longer equal to the mass and a further study is needed to fix their relation. The corresponding equations of motion (of the second and third kind, according to \cite{Infeld:1960}) are more complicated and the validity of the equivalence principle is not guaranteed. For an overview of different approximation methods in the context of the relativistic problem of motion see \cite{Puetzfeld:etal:2015}.

Experimentalists are encouraged to use our results as a framework to systematically test and constrain the effects of scalar fields in gravity. 

\begin{acknowledgments}
D.P.\ was supported by the Deutsche Forschungsgemeinschaft (DFG) through the grant SFB 1128 (geo-Q). D.P.\ thanks S.E.\ Gralla for interesting discussions regarding his approximation scheme.
\end{acknowledgments}
\bigskip

\appendix

\section{Conventions \& Symbols}\label{conventions_app}

Our basic conventions are as in \cite{Puetzfeld:Obukhov:2013:3}. In particular, we use the Latin alphabet to label the spacetime coordinate indices. The Ricci tensor is introduced by $R_{ij} := R_{kij}{}^k$, and the curvature scalar is $R := g^{ij}R_{ij}$. Note that our curvature conventions differ by a sign from those in \cite{Synge:1960,Poisson:etal:2011}. The signature of the spacetime metric is assumed to be $(+1,-1,-1,-1)$, and $\kappa = 8\pi G/c^4$ denotes Einstein's gravitational constant

\section{Brans-Dicke theory}\label{BD}

A wide class of scalar-tensor theories is described by the action $I_{\rm tot} = \int d^4x\,{\stackrel J{\mathcal L}} + I_{\rm m}$ on the spacetime with the metric ${\stackrel Jg}{}_{ij}$. The gravitational Lagrangian density reads ${\stackrel {J}{\mathcal L}} = {\stackrel J {\sqrt{-g}}}\,L({\stackrel Jg})$ with
\begin{eqnarray}\label{LJ}
L({\stackrel Jg}) = {\frac 1{2\kappa}}\biggl(- F^2R({\stackrel Jg}) + {\stackrel Jg}{}^{ij}{\stackrel J\gamma}{}_{AB}\partial_i\varphi^A\partial_j\varphi^B - 2{\stackrel JU}\biggr). \nonumber \\
\end{eqnarray}
This generalizes the Brans-Dicke theory \cite{Brans:Dicke:1961} to the case \cite{Damour:etal:1992} with $N$ scalar fields $\varphi^A$ (capital indices $A,B,C = 1,\dots, N$ label the components of the multiplet). Here  
\begin{equation}\label{AU}
F = F(\varphi^A),\quad {\stackrel JU} = {\stackrel JU}(\varphi^A),\quad {\stackrel J\gamma}_{AB} = {\stackrel J\gamma}{}_{AB}(\varphi^A).
\end{equation}
The action $I_{\rm m} = \int d^4x {\stackrel J {\sqrt{-g}}}\,L_{\rm m}(\psi,\partial\psi,{\stackrel Jg}{}_{ij})$ describes the universal minimal coupling of the matter fields $\psi$ to gravity. 

The metric ${\stackrel Jg}{}_{ij}$ measures distances in the {\it Jordan reference frame} and determines the Riemannian curvature scalar $R({\stackrel Jg})$. Making a conformal transformation
\begin{equation}
{\stackrel Jg}{}_{ij}\longrightarrow g_{ij} = F^2{\stackrel Jg}{}_{ij}\,,\label{ggt} 
\end{equation}
we obtain a different metric on the spacetime manifold. This is called an {\it Einstein reference frame}. 

In the Einstein reference frame the action reads $I_{\rm tot} = \int d^4x\,{\mathcal L} + I_{\rm m}$ where the gravitational Lagrangian density ${\mathcal L} = \sqrt{-g}L$ with
\begin{eqnarray}
L = {\frac 1{2\kappa}}\left(- R + g^{ij}\gamma_{AB}\partial_i\varphi^A\partial_j\varphi^B - 2U\right),\label{LE} 
\end{eqnarray} 
and the matter action
\begin{eqnarray}
I_{\rm m} = \int d^4x\sqrt{-g}F^{-4}L_{\rm m}(\psi,\partial\psi,F^{-2}g_{ij}).\label{LM}
\end{eqnarray}
The scalar curvature $R(g)$ is constructed from the Einstein metric $g_{ij}$, and
\begin{equation}
\gamma_{AB} = {\frac 1{F^2}}({\stackrel J\gamma}{}_{AB} + 6F_{,A}F_{,B}),\qquad 
U = {\frac 1{F^4}}{\stackrel JU}.\label{gamUt}
\end{equation}

\section{Chameleon theory}\label{CH}

In the chameleon theory \cite{Damour:1990,Khoury:PRL:2004,Khoury:PRD:2004}, the universality of the scalar-gravity coupling is abolished. Instead, one assumes that there are several kinds of matter fields $\psi^{(a)}$ that couple to the gravitational field via different metrics $g^{(a)}_{ij} = F^{-2\beta_a}g_{ij}$. The constants $\beta_a$ are different for each kind of matter, and the matter action (\ref{LM}) is generalized to $I_{\rm m} = \sum\limits_a I_{\rm m}^{(a)}$ with
\begin{eqnarray}
I_{\rm m}^{(a)} = \int d^4x\sqrt{-g^{(a)}}L_{\rm m}^{(a)}(\psi^{(a)},\partial\psi^{(a)},g_{ij}^{(a)}).\label{LMC}
\end{eqnarray}
Different matter types $\psi^{(a)}$ do not interact with each other directly. The gravitational action $\int d^4x\,{\mathcal L}$ has the usual form determined by the general Lagrangian (\ref{LE}) of a Brans-Dicke scalar-tensor theory in the Einstein reference frame. 

The coefficients $\beta_a$ appear in the conservation laws of the energy-momentum tensor, which for each kind of matter have the generic form (\ref{cons}).

\bibliographystyle{unsrtnat}
\bibliography{stequiv_bibliography}

\end{document}